\documentclass[fleqn,twoside]{article}
\usepackage{epsf}
\usepackage[cp1251]{inputenc}
\usepackage[russian]{babel}
\usepackage{psfig}
\usepackage{gc}

\heads{Yu.N.Parijskij, N.N.Bursov et al.}
      {RATAN-600 NEW ZENITH FIELD SURVEY AND CMB PROBLEMS}
 
\begin{document} 
\twocolumn[

\Arthead{10}{2004}{4 (40)}{1}{10}
 
\Title {RATAN-600 NEW ZENITH FIELD SURVEY AND CMB PROBLEMS }

 \Authors{Yu.~N.~Parijskij\foom 1, N.~N.~Bursov\foom 1, A.~B.~Berlin\foom 1,
A.~A.~Balanovskij\foom 1, V.~B.~Khaikin \foom 1, \\
E.~K.~Majorova\foom 1,
M.~G.~Mingaliev\foom 1,  N.~A.~Nizhelskij\foom 1, O.~M.~Pylypenko\foom 2, \\
P.~A.~Tsibulev\foom 1,
O.~V.~Verkhodanov\foom 1, G.~V.~Zhekanis\foom 1,
Yu.~K.~Zverev\foom 1}

{Special Astrophysical Observatory of RAS, 369167 Karachaj-Cherkesia, Nizhnij Arkhiz}
{\it ``Saturn'', 252148 Ukraine, Kiev, pr50 Richya Zhowtnya, 2B }

\Abstract{
We present new RATAN-600 data on the synchrotron Galaxy radiation at the
PLANCK Mission and WMAP frequencies at high Galactic latitudes
upto $\ell=3000$. The difference between the standard
synchrotron template ($\ell<50$) of
the WMAP group and RATAN-600 data was detected
with the strong synchrotron ``longitude quadrant asymmetry''.
It may change the
WMAP estimates of $z_{reheating}$ from low $\ell$ polarization data.
The polarized synchrotron noise for very deep
observations ($\ll1\,\mu$K) at the PLANCK HFI
was not detected at $\ell>200$ scales.
``Sakharov Oscillations'' in the
E-mode ($500<\ell<2000$) should be well visible even at $\sim $10GHz.
The polarized noise from relic gravitational waves ($\ell\sim80$) may be
confused with B-mode of synchrotron
Galaxy polarized noise at the frequencies below 100\,GHz, but there are no
problems at HFI\-band.
}

]
\email 1 {par@sao.ru}

\section{Introduction}

The synchrotron and cosmology synchrotron noise from the Galaxy is one of the
background screens between the early Universe and the observer. But for
polarization experiments this screen may be the most dangerous due to
possible high and frequency-- dependant E-- and B--modes of polarization
(up to 70{\%}).

It is not easy to extrapolate available nice maps of the Galaxy
synchrotron emission from decemeter low resolution data to PLANCK
frequency and to the  scales important for Cosmology. The first problem-
unknown variations of spectral index with frequency and space, the second-
correction for Faraday effect.

The first broad review of the problem was done by M.Tegmark (Tegmark
et al., 1999), with estimation of the range of possible effects in the
Cosmology important part on the frequency- scale plane. The so cold
"Pessimistic", "Middle", and "Optimistic" variants were suggested.

Just after this paper we began to accumulate data on the Galaxy background
with RATAN-600 multi-frequency receivers array ($\sim $30 channels in
the 0.6GHz - 30 GHz band in I, L, R, U, Q Stocks parameters and with
different resolution from few arc seconds to few arc minutes.

Some preliminary results have been already published (Parijskij,
2000, 2003,  Parijskij and Berlin, 2002; Parijskij and Bursov, 2002;
Parijskij and Novikov, 2004). They were connected
with spinning dust problem and new limit was found for this screen,
much below Tegmark "Pessimistic" case at least at $\ell=1000$, most important
for CMB E-mode of polarization. New limit was also found  for magnetic
dust polarization, suggested by Princeton group recently. "Faraday"
Galaxy noise was checked at LFI band. This noise can destroy the purity
of the theoretical $<B>=0$ requirements for Thomson scattering.

Several recent experiments demonstrated, that synchrotron Galaxy noise
has to be studied deeper, than before by several reasons (Naselsky et
al., 2003)

 1. The unexpectedly high Thomson scattering between the
 recombination
epoch and observer, deduced from very strong polarization at low $\ell$. It
contradicts  Ly-breaks results and requires new population of z$ \gg $6
objects for early ionization of the Universe. Several alternative
interpretations appeared in literature, and Galaxy polarization is in
this list.

 2. Strong interest in the processes of z=1000 recombination increases
the importance of the polarization measurements of "Sakharov Oscillations"
and many groups are waiting for much better information on the Galaxy
polarization data.

 3.The fundamental check ("experiment cruces") of the Inflation
scenario- discovery of relic grav. waves. B-mode polarization at
$\ell\sim 100$ was suggested as the direct indication of the existence of
the primordial grav. waves, (Zaldarriaga, 1995) predicted by I.Novikov
in 60-ties. The reality of this experiment depends  on the power of
Synchrotron noise (which has $<B> = <E>$, contrary to scalar Thomson
effect, with $<B>=0$) at $50<l<1004$ band.

4. The primordial magnetic field may be traced through polarization
measurements by Faraday effect at z=1000 and by detection of the large
(larger than z=1000 horizon) polarization scale-inflation induced
magnetic field. Again, Galaxy polarization limits the accuracy of such,
sub-micro K effects at $30<l<200$ scales. This limit well we do not
know  yet.

 5. After "ARHEOPS" experiment, with detection of strong polarization
from the Galaxy dust screen at HFI band, we have to find how far we can
go to the LFI band, to be in the best place, between Synchrotron and
Dust noise ("Scylla, $f^{-3}$, and Haribda, $f^2$," ~situation). The better
information on the synchrotron, the better estimates of effects from
other screens, including the most uncertain polarized "spinning dust" one.

With our 600m- aperture reflector, RATAN-600, we have no limitation in
angular resolution even at 1 GHz at all scales important for Cosmology
and all $\ell>\ell_{max}$ for PLANCK HFI may be checked on the synchrotron by
observations at $f \ll f_{PLANCK}$. To be as deep in pixel sensitivity as
possible, we selected limited portion of the sky
($f_{sky}=\Omega / 4\pi=0.01$)
and exposed each pixel in this field as long as $\sim $1 Day pixel at
$\ell=200$ scales. It may be compared with $\sim $ 200 seconds for PLANCK
mission 2007 and 47 seconds for WMAP.  $f^{-3}$ factor solves the sensitivity
problem. Indeed, even 10mK data at 1GHz correspond to 10 nano-K at 100GHz,
main HFI frequency.

Full data will be presented to PLANCK consortium, but the most important
results will be given here.

\section {Observations}

All RATAN-600 (Parijskij, 2003) observations were done in the standard
one-sector mode (North sector) with the secondary mirror N1. This mirror
(parabolic cylinder) was equipped by multi-frequencies receivers
array, installed along the focal line. In the standard transit mode of
observations, the source image is moving along this line, and in $\sim
$1 minute frequency spectrum of Sky of the beam size at all 31 channels
appears at the common backend.  0.6 GHz, 1 GHz, 2.4 GHz bands were
divided into 8 independent channels; 3.9 GHz, 7.7 GHz, 12 GHz and 21.8 GHz
had cryo-HEMT receivers. At the 30GHz we used 6- feeds matrix HEMT
receivers with MMIC technology, and CMB polarization in the Stocks Q
parameter were accumulated. All $f>2.4$ GHz receivers had few mK NET, with
best sensitivity at 3.9GHz ($\sim $2mKs$^1/2$). All receivers at $f<3.9$ GHz
had 10-30mK NET ({\tt http://www.sao.ru}; Parijskij and Korolkov,
1986; Berlin and et al., 2000).

Local zenith field was selected by several reasons. At zenith instrumental
aberrations do not exist and (Stotsky, 1972)
there are limitation in the size of
the receivers array. At zenith the orientation of the main surface
panels is correct, and no ``Diagonal errors'' appear (Braude et al.,
1972). At 45 degrees inclination, the random panel errors are less by
$\sqrt{2}$. During the panelsurface adjustments,
the radius of panels has been
optimazed to reduce the ``panel curvature error'' in the zenith mode.

\begin{figure}[!t]
\fbox{\psfig{figure=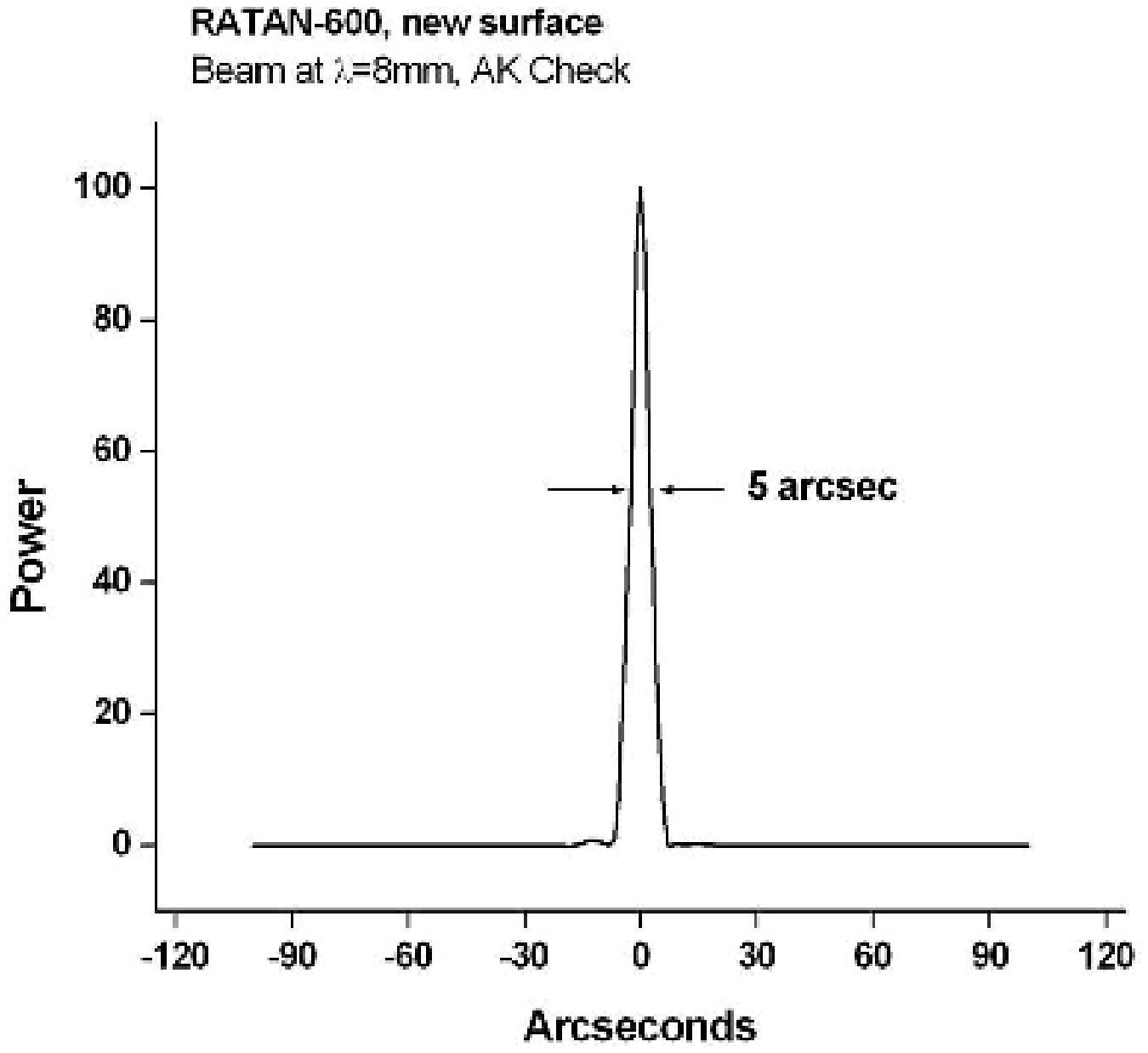,width=8cm}}
{Fig\,1.
The shape of the one-sector beam at the highest frequency, 32\,GHz,
after re-surfacing of all panels and with  new accurate panel adjustments
system. The panel r.m.s. error was improved by factor 5, that reduced
the wide angle scattering at 1\,cm
}
\end{figure}

The transit time of the $\ell=200$ scale at zenith was about 500 seconds, and
sub-mK white noise pixel sensitivity may be expected at all frequencies
in the single day transit. The white noise component at all frequencies
in the accumulated data was much below 1mK. For the synchrotron noise,
1mK at 7.6\,cm corresponds to 20 nano-K at PLANCK frequency 150 GHz, and
$20 \mu$K at WMAP frequency 23 GHz, and we were not limited by white
noise at $\ell<200$.

\begin{figure}[!t]
\fbox{\psfig{figure=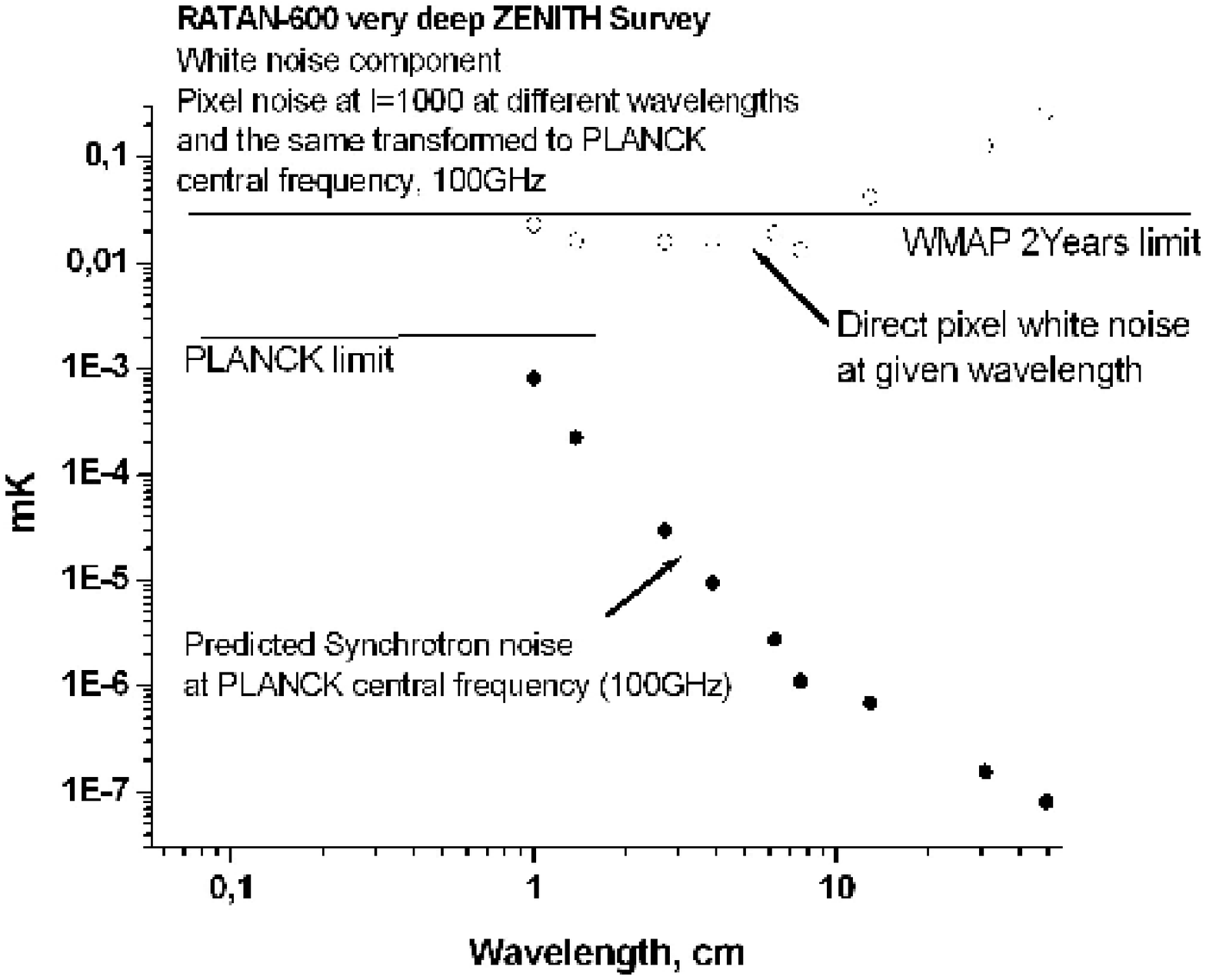,width=8cm}}
{Fig\,2.
It demonstrates, that white noise
really does not limit the accuracy of detection Galaxy Synchrotron noise,
if we extrapolate it to PLANCK HFI.
}
\end{figure}

The $1/f$ noise and interference are the real limitations. The sum and
difference between 2 groups of observations were used to find their
effects.  Up to 500 24-hours daily scans were carefully analyzed, and
$C_\ell$ - structure of the real noise was estimated.
At high $\ell$ this noise is
below white noise component, at $\ell>200$ it dominates at all frequencies
with the slope in the FFT of the 14-scans close to -1. The standard rule
for Galaxy screens- $C_\ell \sim \ell^{-3}$,
and at very big scales these screens
dominate at all frequencies.

The conversion from $T_a$ to $T_b$ for given pixel at given
$\ell$ needs the beam
de-convolution. WMAP results are too noisy for reconstruction of the $T_b$
maps at $\ell>50$. We decided to use theory at high $\ell$. To simplify the task,
we realized CMB model of the sky, using standard HEALPIX algorithm up
to $\ell=3000$, convolved it with new version of RATAN-600 beam (Majorova, 2000)
to simulate
24-hours transit scan with different 2D- beams from 30GHz to 0.6GHz
and compared the dispersion of convolved data with the not convolved
one. Simple dispersion analysis does not show very strong effect,
it is due to dominant role of low  $\ell$ CMB anisotropy noise in the
dispersion. From the FFT of the convolved and not -- convolved sky map we
can find the correction factor ($T_{a} / T_{b}$ ratio) for any given scales
or $\ell$.

As we expected, the convolution effect is small at high frequencies,
medium at low frequencies and low $\ell$, but very strong at high $\ell$ and
low frequencies.


\begin{center}
\begin{tabular}{|r|c|c|c|c|c|} \hline
 wave, cm & $\ell$~1000 & $\ell$~200 & $\ell$~80 & $\ell$~2-10  \\
\hline
 1    &  1   & 1   & 1   &1 \\ 1.4  &  1   & 1   & 1   &1 \\ 2.7  &
 0.95 & 1   & 1   &1 \\ 3.9  & 0.8  &0.85 &0.9  &1 \\ 7.6  & 0.36 &0.77
 &0.8  &1 \\ 13   &0.154 &0.58 &0.66 &1 \\ 31   &0.05  &0.38 &0.58 &1 \\
 49   &0.01  &0.26 &0.53 &1 \\
\hline
\end{tabular}
\end{center}

We can compare these BEAM losses  with WMAP and PLANCK losses. For $\ell=1000$,
the most important scale for CMB polarization, at the highest resolution
they are about 0.001(WMAP) and 0.05(PLANCK).

To estimate Synchrotron noise at PLANCK frequencies, say, at 3 mm, we
should take into account not only the $[f_{PLANCK} / f_{RATAN-600}]^3$ factor,
but also confusion noise, receivers and $1/f$ noises, and BEAM losses.

The closer we are to the NVSS (FIRST) frequency, 1.4GHz, the deeper
background sources cleaning may be done. In contrast, strong interference
(and by factor 3-10 more receivers noise) at $f<$3GHz, great BEAM losses,
prevents to  realize frequency cube- factor. We have found, that
7.6-31 cm band has the main priority.  The main result at present- is the
independent data on the synchrotron Galaxy noise at different frequencies
and scales.

Below we show  some decimeter results with their sums and differences
at all $\ell$, in the band, important for polarization on relic grav. waves
($\ell \sim 80$ (Zaldarriaga, 1995)), and at first Doppler peak, $\ell=200$.

At $\ell1000$ and $\ell2500$ we used 7.6cm, but $\ell^{-3}$
extrapolation from
decimeters gives comparable result, in spite of big beam losses. (We
should remind, that for RATAN-600 $\ell_{max}=D / \lambda$, which is much
greater, than ${S_{eff}}^{-1/2}/\lambda$. Also, in standard approximations
of the BEAM by Gaussian shape, the physically inescapable $\ell_{cut-off}$
($u,v_{cut-off}$ equivalent)theorem is ignoring.
For RATAN-600 $\ell_{cut-off}$ at the
lowest frequency, 0.6GHz, is equal to $2\pi*u_{cut-off}=2\pi*D/\lambda=5000$.)

\section {Discussion}

We have tried to compare our deep data at decimeters with WMAP synchrotron
template. We convolved WMAP synchrotron sky with RATAN-600 beams at all
frequencies and compared the simulated scans with the real one. We found
strong difference between WMAP template and all our scans.

This difference may be interpreted as the wrong spectral index variations
template across the sky. This large scale difference may result in the wrong
interpretation of the large scale polarization WMAP data (see discussion
in (Naselsky, 2003)) at low $\ell$ ($2<\ell<8$).

Great Thomson depth may be not the only interpretation -- Galaxy
synchrotron polarization is another one.  It is not possible to compare
our results with WMAP at high $\ell$ (in WMAP data signal-to-noise ratio $<1$ at
$\ell>50$), and we used our own new data here.  The main result is that there
are  no problems with synchrotron polarization at the most interesting
for present day experiments at $\ell=1000$ at frequencies above 10-20 GHz.

\begin{figure}[!t]
\fbox{\psfig{figure=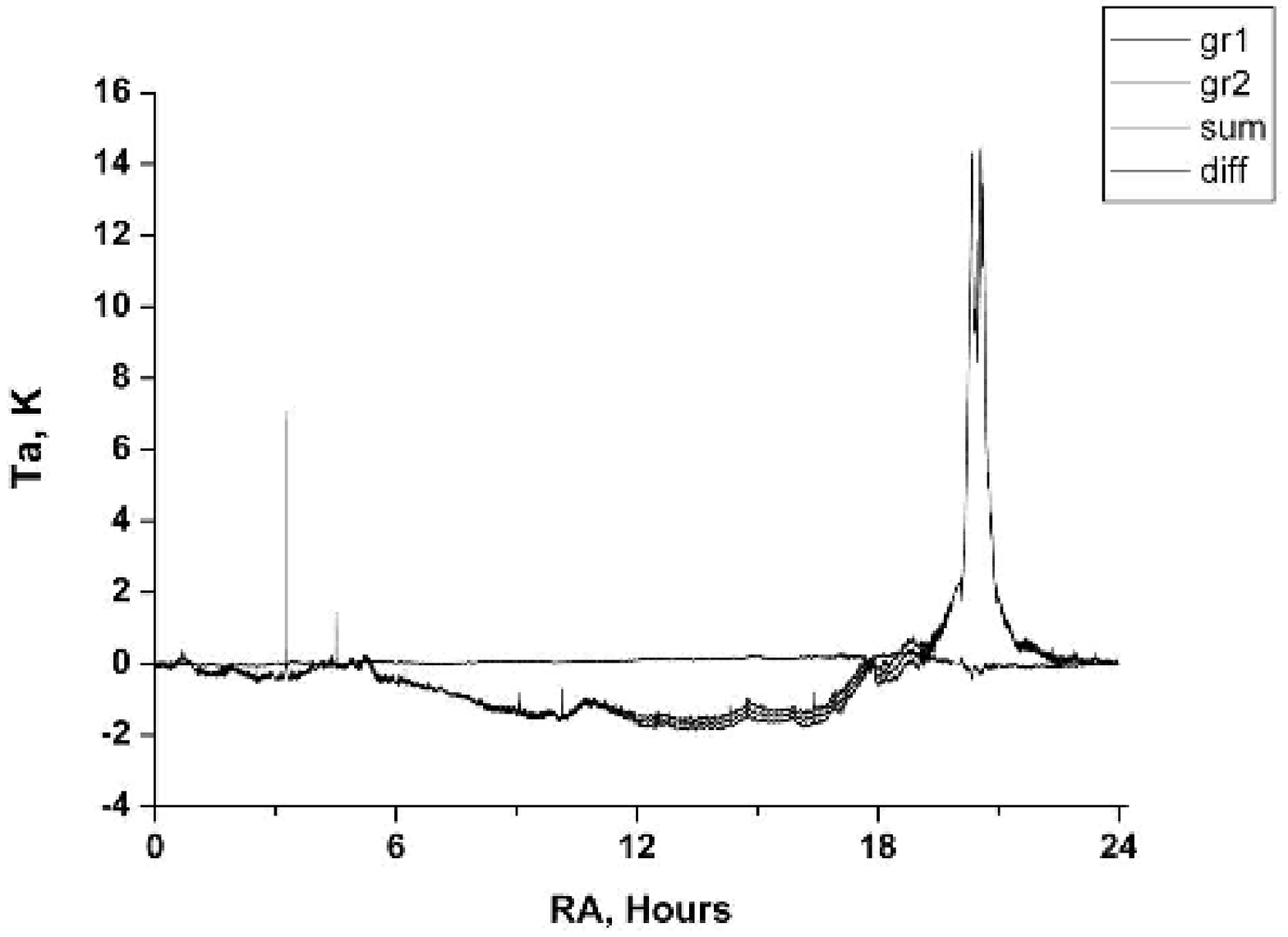,width=8cm}}
{Fig\,3.
1GHz result. Two independent groups,
gr1 and gr2, of observations with mean (``sum'') and difference (``diff'')
between groups. About 300 24- hours scans were used. Milky Way dominates
at $5^h$ and $20^h$, the minimum is close to the bII maximum region, $12^h$.
It
is the deepest Galaxy cross-cut at high bII (2mK temperature resolution)
at high $\ell$ scales (up to $\ell=6000$).
}
\end{figure}

\begin{figure}[!t]
\fbox{\psfig{figure=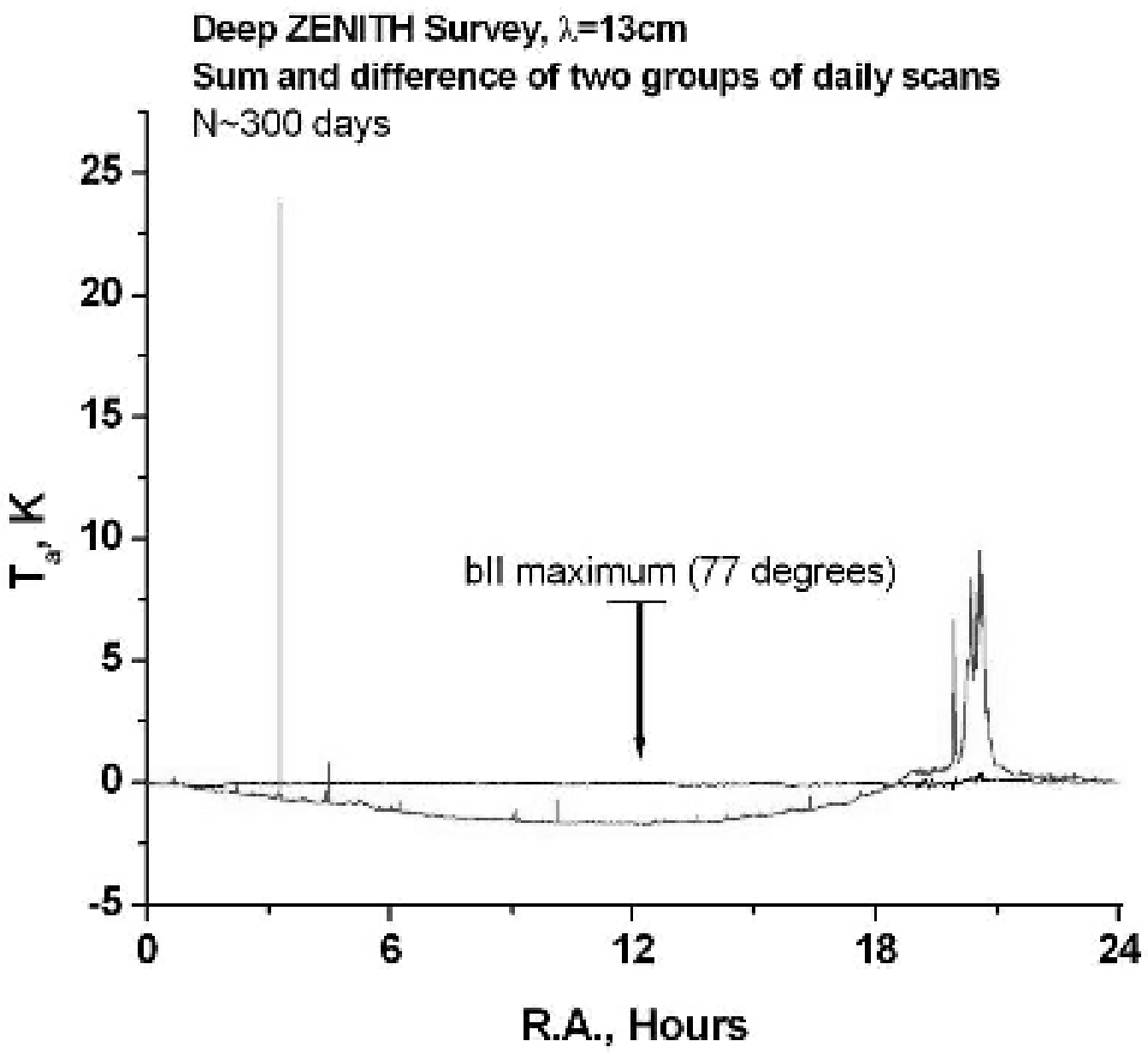,width=8cm}}
{Fig\,4.
Very deep confusion limited cross-cut at 13 cm. 300 daily scans wave used.
Note, that at this wavelength the coldest sky is close to the bII
maximum point.
}
\end{figure}

\begin{figure}[!t]
\fbox{\psfig{figure=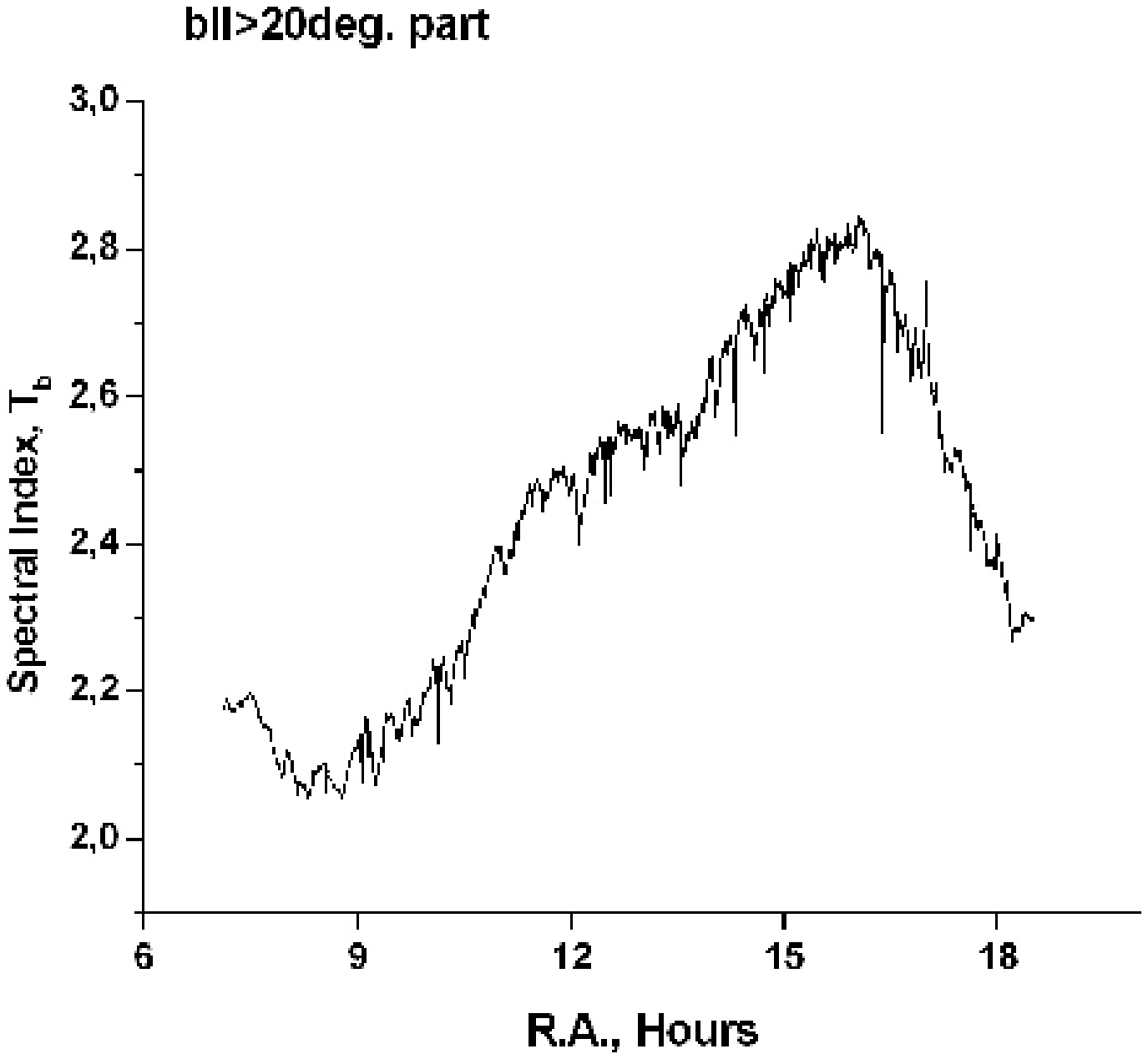,width=8cm}}
{Fig\,5.
Variations of the Galaxy spectral
index across the strip at bII > $20^{\circ}$ Haslam 73 cm and RATAN-600
7.6 cm date wave used. bII is maximum at R.A. $\sim    13^h $,
but it is not the coldest part of sky.
}
\end{figure}

\begin{figure}[!t]
\fbox{\psfig{figure=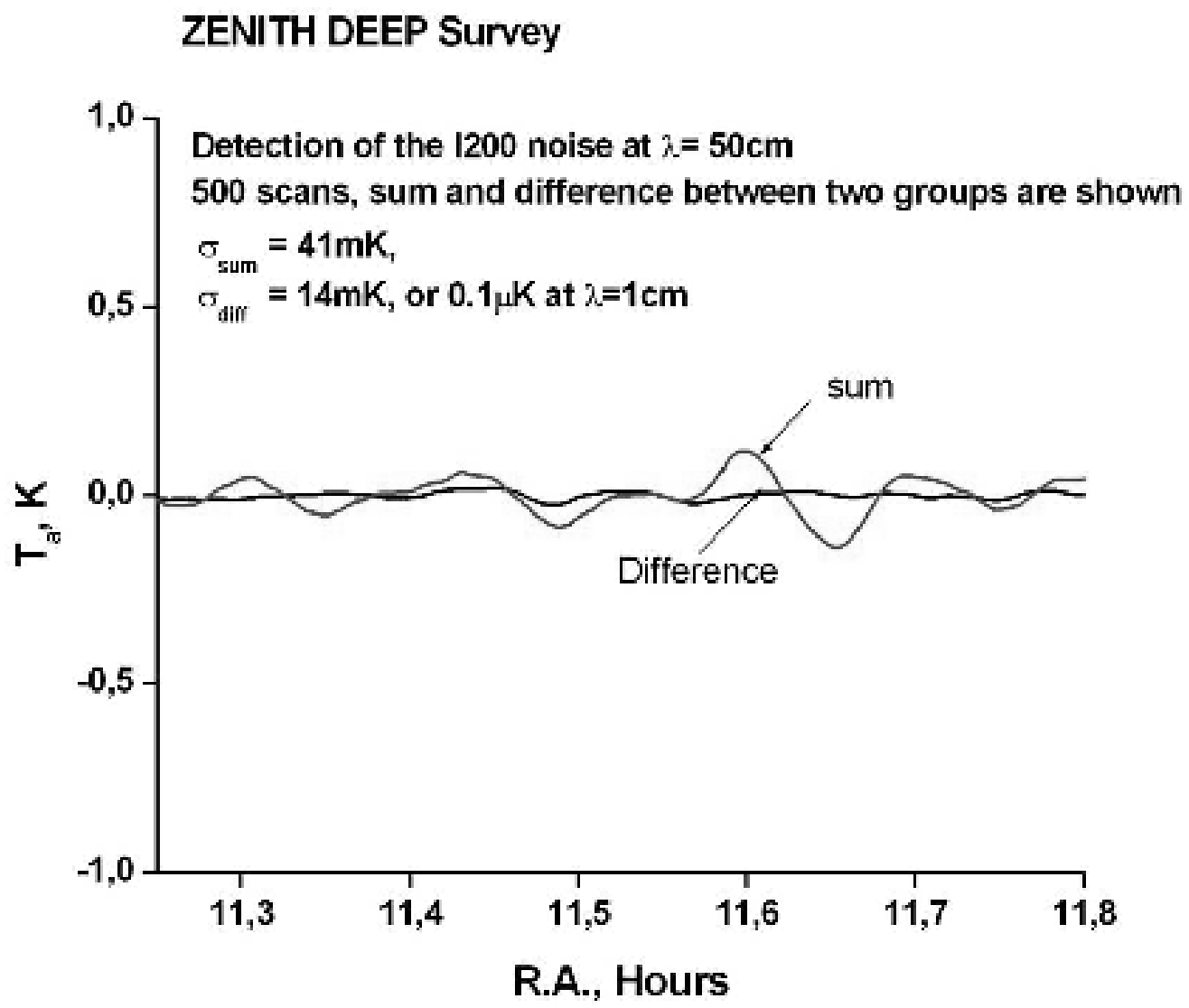,width=8cm}}
{Fig\,6.
First detection of the Synchrotron
noise at 0.6 GHz (50cm) , $\ell=200$ scale.
The rest of confusion noise is also
here, and we suggest 41mK as the upper limit (4mK-8mK for polarization)
}
\end{figure}

At the most important for relic grav. waves polarization, $80<l<100$,
(Zaldarriaga, 1995) much higher frequencies have to be used, but at the
central HFI band synchrotron Galaxy noise is below the relic grav. waves
noise.

Relic grav. wave experiments are based on the theory, which predicts
$0.1 \mu$K effects, and belongs to the third generation anisotropy
experiments, after ``Sakharov Oscillations'' (10-100 $\mu$K), E-mode
polarization (few $\mu$K).  Spectral features in the CMB anisotropy are the
only one field, where effects may by order of magnitudes weaker ($0.01 \mu$K).

These next generation experiments need in very deep investigations of all
screens, involved in observations. Synchrotron space- frequency variations
are one of the difficult problems at least in the low frequency band.

\begin{figure}[!t]
\fbox{\psfig{figure=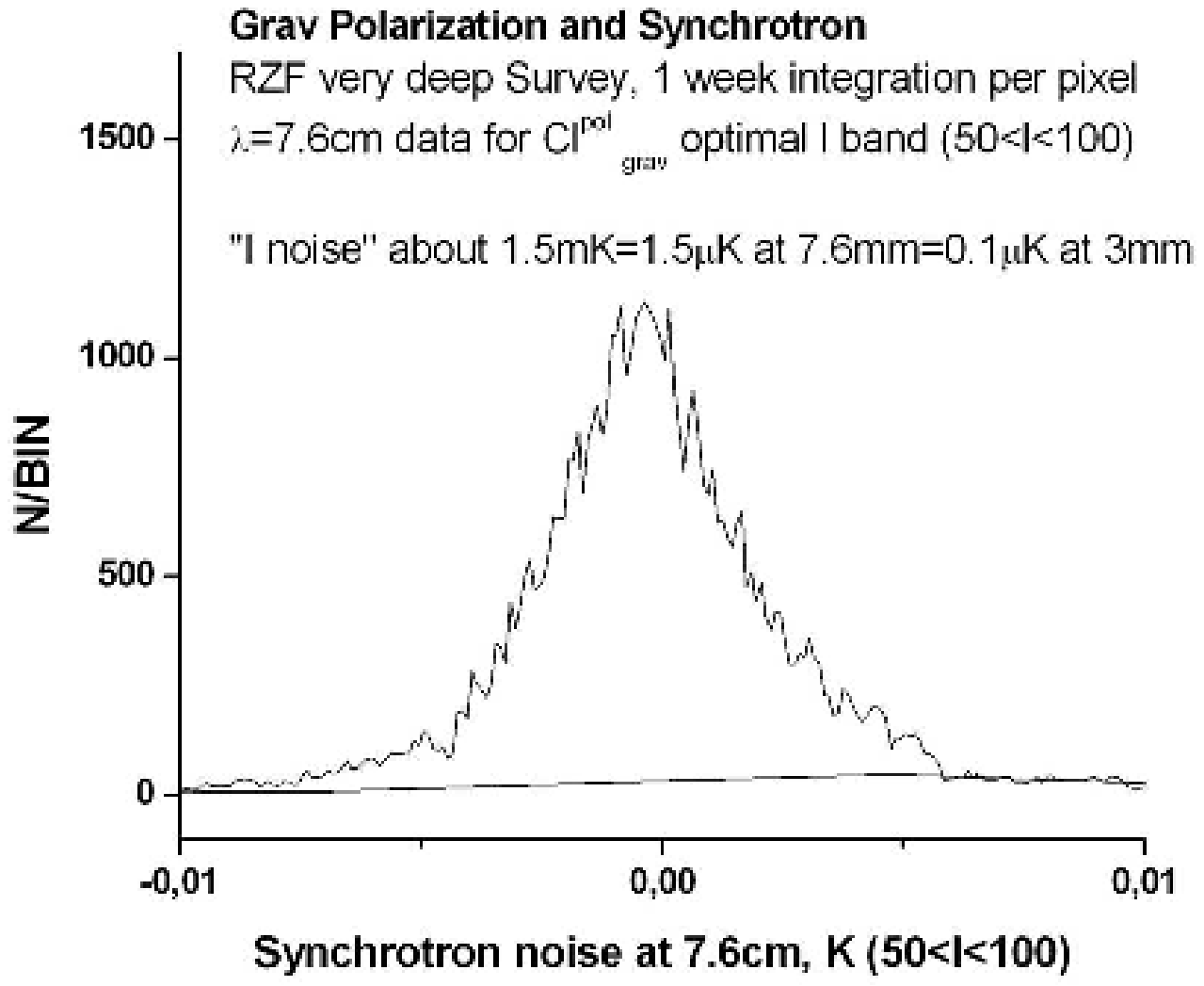,width=8cm}}
{Fig\,7.
Relic Grav. Waves polarization and
Galaxy Synchrotron noise at the best $\ell$, suggested by theory (Zaldarriaga,
1995) $\ell \sim 80$. In this band we have 1-2 mK noise in I Stocks parameter
(as an upper limit); Even with 50{\%} polarization, it corresponds to
less than $0.1 \mu$K at PLANCK HFI.
}
\end{figure}

\begin{figure}[!t]
\fbox{\psfig{figure=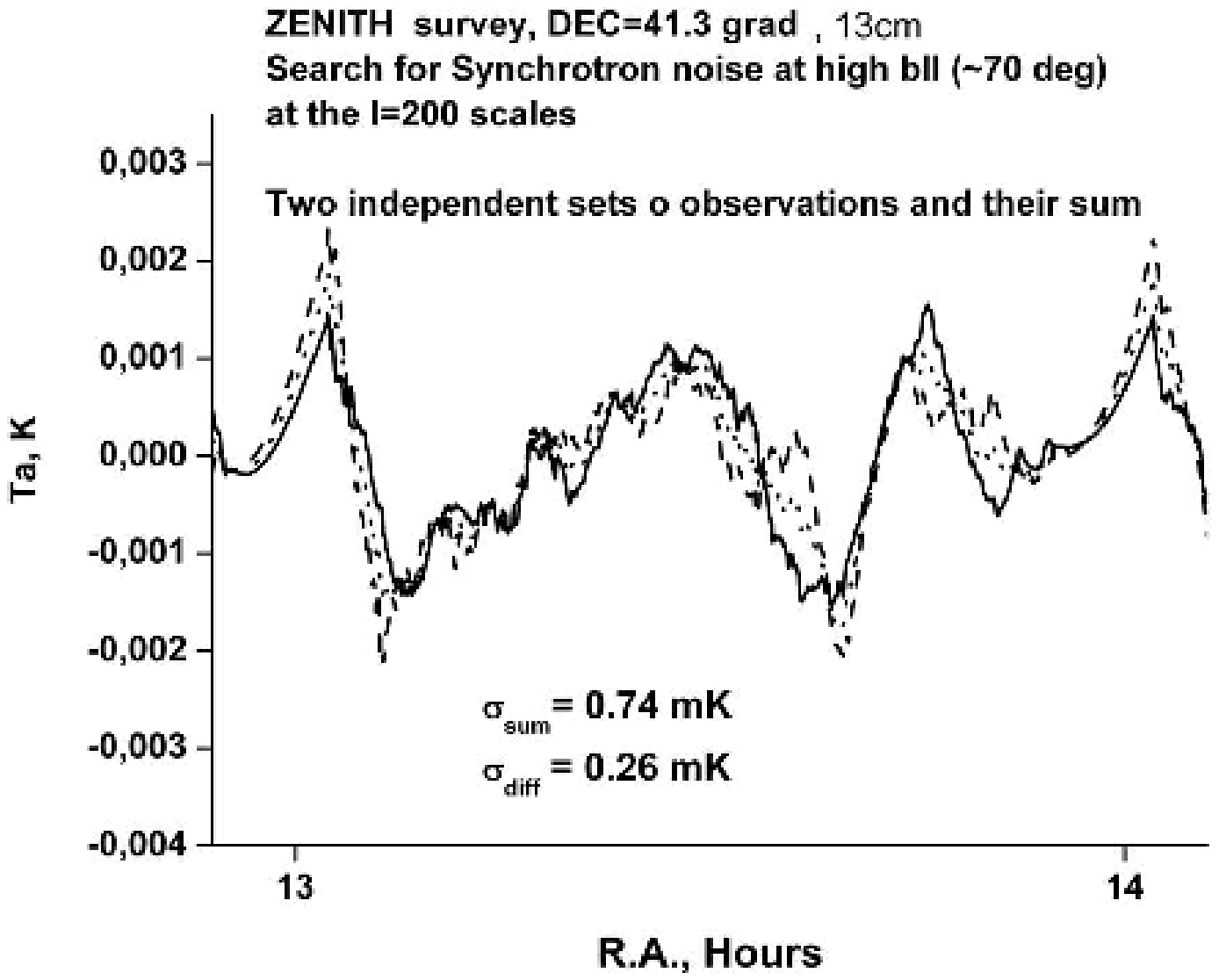,width=8cm}}
{Fig\,8.
13cm. (2.4GHz). Real Galaxy
synchrotron temperature variations at $100<l<300$ scales, but with sub-mK
amplitude (about 10 nano-K) at PLANCK HFI.
}
\end{figure}

\begin{figure}[!t]
\fbox{\psfig{figure=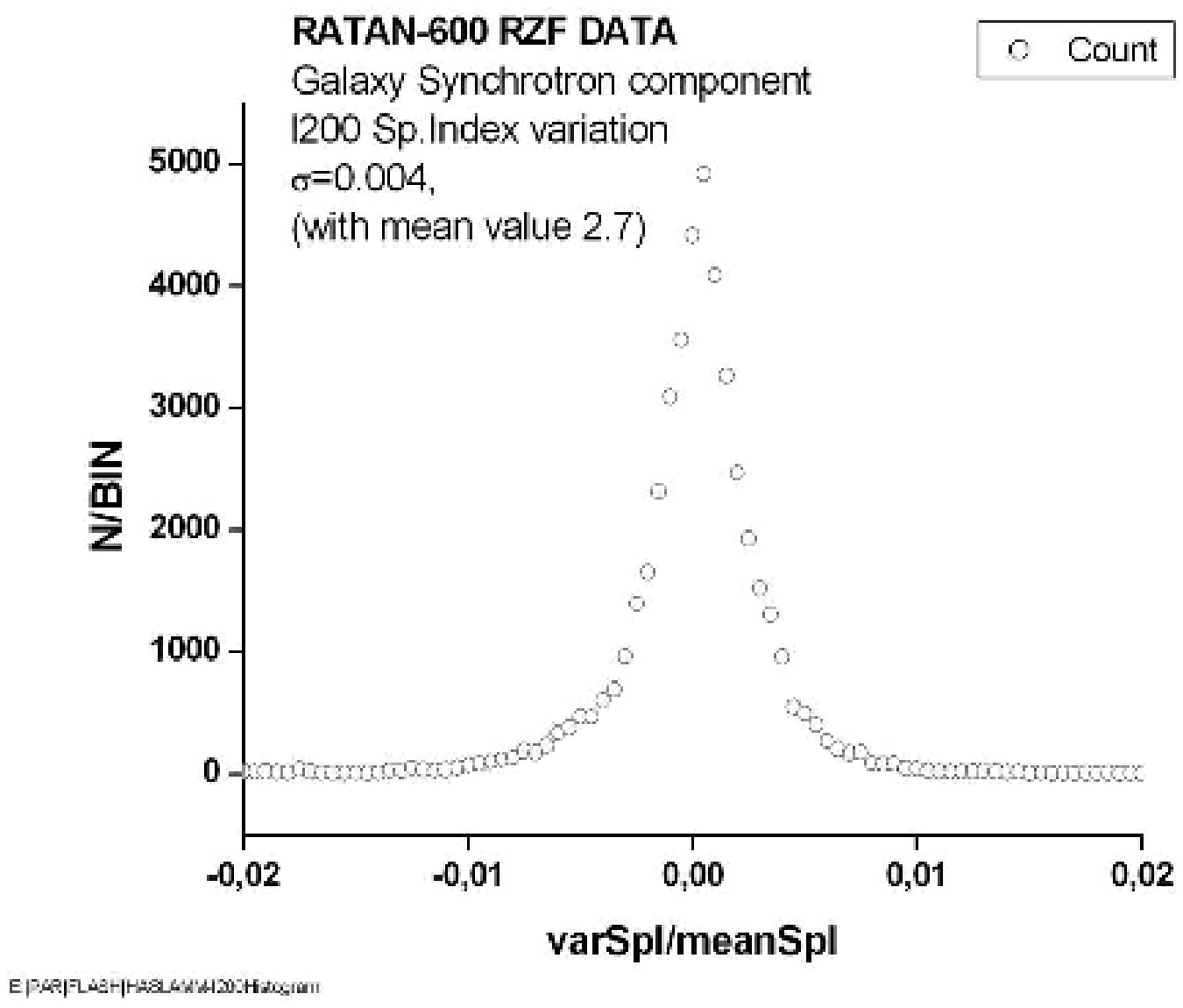,width=8cm}}
{Fig\,9.
Galaxy Synchrotron noise at the main "Sakharov Oscillations" peak,
$\ell=200$. Very small variations of the $T_b$ spectral index were found. This
``horizon scale'' spectral-space variations limits the accuracy of the
CMB spectroscopy.
}
\end{figure}

\begin{figure}[!t]
\fbox{\psfig{figure=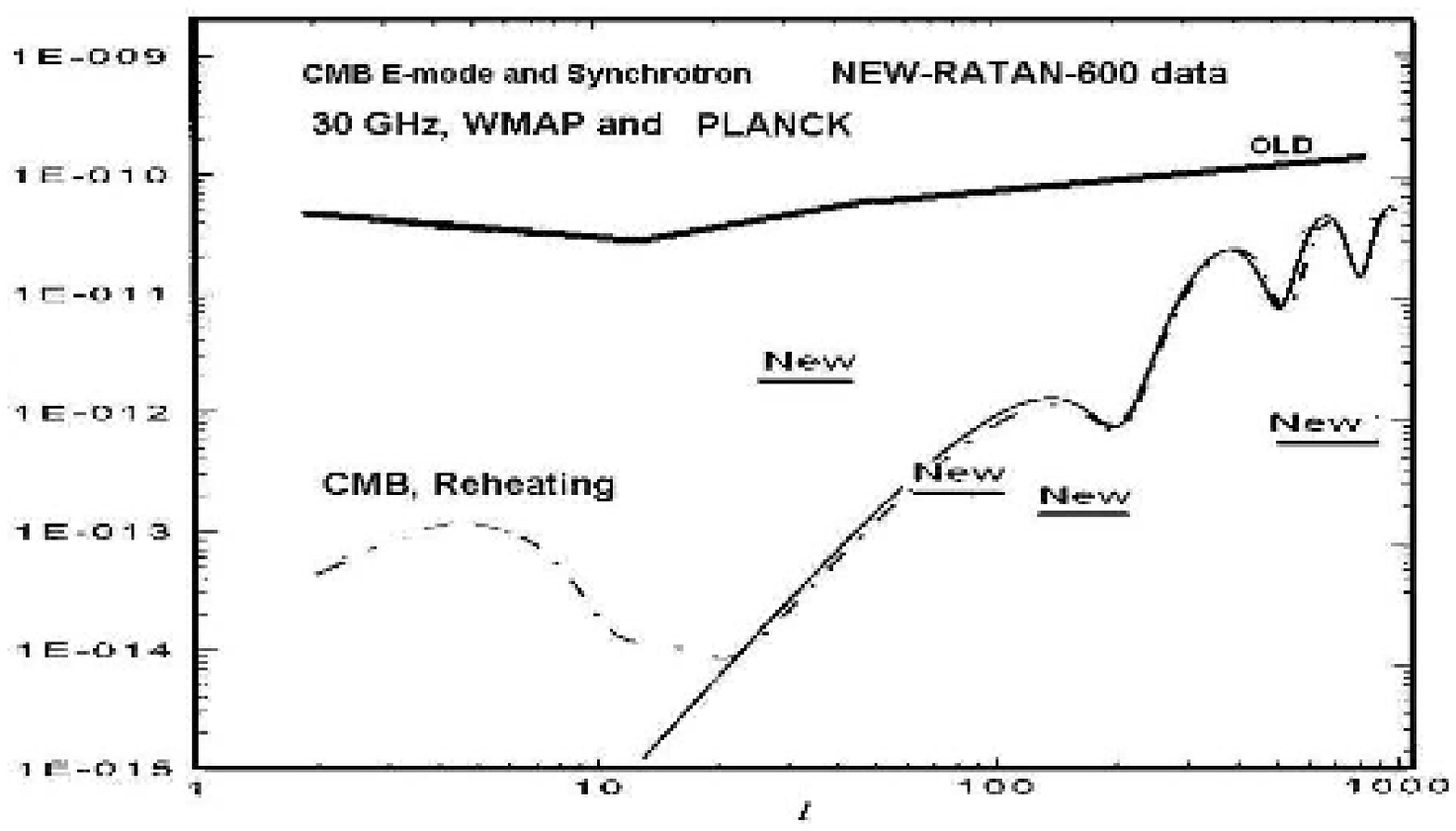,width=8cm}}
{Fig\,10.
E-mode of "Sakharov Oscillations"
and new RATAN-600 data on the upper limit of Galaxy synchrotron
polarization noise. We extrapolated noise from decimeter RATAN-600 data to
30GHz with better spectral index information. New data are superimposed
on the recent world collection of the synchrotron polarization at
different scales at 30GHz (Cortiglion  et. al, 2004). In contrast to
(Cortiglion  et. al, 2004), no problem exist at $\ell=1000$ at this PLANCK
LFI frequency band, as well as at ``Cosmological Gene'' main frequency
(30\,GHz). Problems may appear at 10GHz and below.
}
\end{figure}

The more accurate are synchrotron data, the better estimations of the
others screens may be done. Free-free Galaxy noise may be estimated
with accuracy above $H_{\alpha}$ results, which also the subject of dust
and temperature space variation. Free-free screen is not (or has very
small) polarized, but spinning dust screen can be polarized strongly,
and we are going to estimate it effect very soon. Our preliminary data
at $\ell=1000$ where optimistic one (Parijskij, 2003), but now we have
spinnius dust much
deeper data just at the most active frequencies (20-30GHz, 8GHz).

We stopped now the single-DEC data accumulation when $C_l$(signal) began
greater, than $C_l$(real noise) and change the strategy to the multi- DEC
mode, to reduce the "Cosmic Variance" error and effect of "l-m" confusion
(Naselsky et al., 2003).  Now we have a more than 500 sq. deg. field
with central DEC at the position of 3C84.

\begin{figure}[!t]
\fbox{\psfig{figure=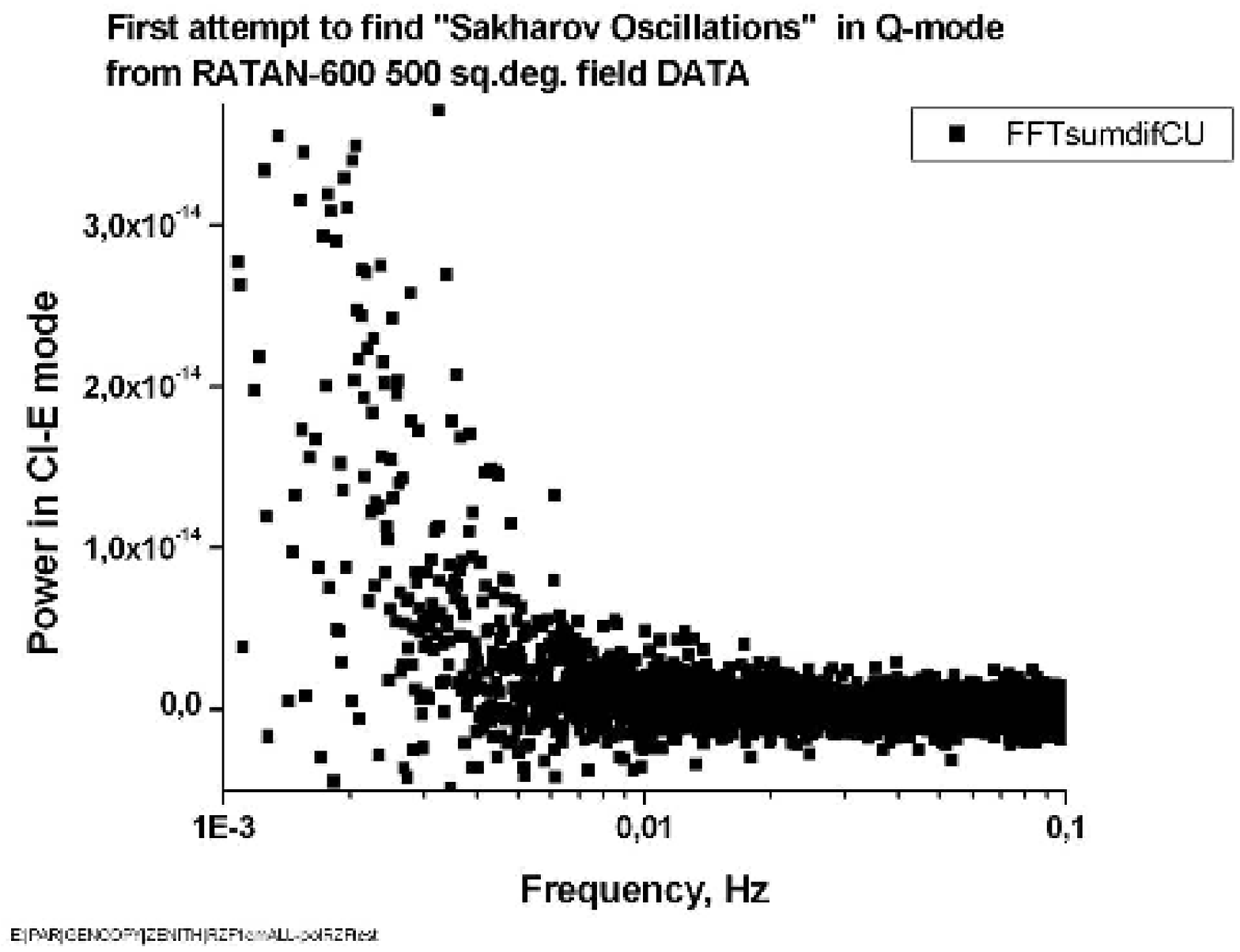,width=8cm}}
{Fig\,11.
30GHz sky polarization data. FFT
of the sum of 500 24-hours scans at 10 adjacent declinations, 12 arcmin
apart in 500-degrees area.  Strong unidentified polarized signal visible
at $\ell < 100$ up to the very low $\ell$.  It is too strong for synchrotron, and
spinning dust and some systematics are under suspicion.
}
\end{figure}

The results presented here are the part of the ``Cosmological Gene
PROJECT'' (Parijskij, 2003; {\tt http://www.sao.ru}; Naselsky et
al., 1999).

\Acknow This work was done with partial support by RBR (grant
02-02-17819), Astronomy (1.2.2.4) and by special grant of Saint-Petersburg
Acad. Center.

\end{document}